\documentclass[aps,prl,reprint,amsmath,amssymb,superscriptaddress,nofootinbib,raggedbottom]{revtex4-2}
\usepackage{bm}
\usepackage{booktabs,array}
\usepackage{float}

\usepackage{xcolor}
\definecolor{darkblue}{RGB}{0,70,140}

\usepackage[
  colorlinks=true,
  linkcolor=darkblue,
  citecolor=darkblue,
  urlcolor=darkblue
]{hyperref}

\hypersetup{pdftitle={The Double Copy of the Doubly Spinning Black Ring},pdfauthor={Damien A. Easson}}
\newcommand{\W}{\mathcal W}
\newcommand{\dd}{\mathrm d}
\newcommand{\Ht}{H_{\rm top}}

\begin{document}
\title{The Double Copy of the Doubly Spinning Black Ring}
\author{Damien A. Easson}
\email{easson@asu.edu}
\affiliation{Department of Physics, Arizona State University, Tempe, Arizona 85287, USA}
\begin{abstract}
We establish an exact curved-background Weyl double copy for the
doubly spinning Pomeransky--Sen'kov black ring. Its full Weyl tensor
is a quadratic form in two source-free Maxwell fields: one generated
by the stationary Killing vector, the other a flux-carrying magnetic
field obtained by linearizing the dipole-ring gauge potential about
its vacuum limit. The mixed Maxwell bilinear is essential: no
spacetime-independent real or complex change of field basis
removes it. We identify the inverse $S_{IJ}$ of the
Weyl coefficient matrix as a matrix zeroth copy. The same matrix arises independently from a weighted
limit of the inner products of two Killing vectors in a
six-dimensional Ricci-flat parent and satisfies
$\Box_g S_{IJ}=-\tfrac12 F_{I\,ab}F_J{}^{ab}$. The singly spinning limit recovers
the diagonal two-field Weyl double copy of the Emparan--Reall ring.
\end{abstract}
\maketitle

\emph{Introduction.---}
In four dimensions, stationary, asymptotically flat vacuum black
holes have spherical horizons and are constrained by Kerr uniqueness.
In five dimensions, by contrast, vacuum gravity admits asymptotically
flat black rings with $S^1\times S^2$ horizons, rotating around
the circle and about an azimuth of the sphere
\cite{EmparanReall2002,PomeranskySenkov2006,EmparanReallReview2006}.
The second angular momentum leaves the horizon topology unchanged
but introduces new structure into the Weyl curvature.

The Weyl double copy expresses gravitational tidal curvature through
quadratic combinations of gauge-field strengths
\cite{Luna2019,MonteiroOConnellWhite2014}.
Tensorial Weyl doubling on the physical curved spacetime has produced
higher-dimensional examples, including the singly rotating
Myers--Perry geometry
\cite{WeylDoubling2020,ChawlaKeeler2023}; further five-dimensional
constructions treat restricted type-D and type-N solutions
\cite{ZhaoD2025,ZhaoN2025}. Related work includes a perturbative type-II fluid double copy
\cite{KeelerMonga2025} and a general metric-level approach
developed by Kent and Zimmerman, including a flat-background
type-I example \cite{KentZimmerman2025}. Sums and mixed products of
field strengths occur in multicopy constructions
\cite{Chacon:2021wbr,BrownSpence2026} and sourced Weyl double copies
\cite{Easson:2021asd,Easson:2022zoh,ArmstrongWilliams2025}.

We seek a Weyl double copy of the doubly spinning ring using
two source-free Maxwell fields. Their squares and mixed product
give three quadratic Weyl tensors. Since constant linear
combinations of Maxwell solutions remain solutions, a change
of field basis can alter the coefficients of these three terms.

The singly spinning ring does not admit a flat-background
Kerr--Schild double copy \cite{Bah:2019sda},
but admits a diagonal two-field Weyl copy
\cite{EassonER2026}. We show that stationary and flux-carrying Maxwell fields
reconstruct the full Weyl tensor of the doubly spinning ring,
but no spacetime-independent change of field basis eliminates
their mixed term.

\emph{Geometry and Maxwell fields.---}
Let $g=g_{\rm PS}$ be the vacuum Pomeransky--Sen'kov metric,
written in the coordinates used by Chen, Hong, and Teo (CHT)
for the dipole-ring family \cite{ChenHongTeo2012}.
The coordinates are $(t,x,y,\phi,\psi)$, both angles have period
$2\pi$, and $\xi=\partial_t$ is normalized by $\xi^2\to-1$ at infinity.
The regular parameter interior and open exterior chart are
\begin{equation}
\begin{gathered}
0<c<1,\qquad 0<b<\frac{1-c}{1+c},\qquad \kappa>0,\\
-1<x<1,\qquad -1/c<y<-1 .
\end{gathered}
\label{eq:domain}
\end{equation}
Here $c$ is a dimensionless ring parameter, with the outer
horizon at $y=-1/c$; $\kappa$ sets the scale, and $b$
controls the second rotation. The limit $b\to0$ gives
the balanced singly spinning ring. Full metric formulae
and the map to standard PS parameters are given in the Supplemental
Material \cite{Supplement}. The regular exterior is described in the
PS global-structure analysis
\cite{ChruscielCortierGarcia2010}.

The first Maxwell field is the stationary Papapetrou field, with
$\xi^\flat$ the one-form dual to $\xi$:
\begin{equation}
E=\dd\xi^\flat,\qquad E_{ab}=2\nabla_{[a}\xi_{b]} .
\label{eq:E}
\end{equation}
Its closure and source-free Maxwell equation follow from the Killing
identity in vacuum. Killing-generated electromagnetic test fields on
black rings have been studied previously \cite{OrtaggioPravda2006}, and
Killing fields and geometric scalar seeds have appeared in other
classical double-copy constructions
\cite{EassonIsometries2023,Easson:2026xod}.
To generate the flux-carrying field, we introduce a signed amplitude
$\epsilon$ into the dipole family \cite{EmparanDipole2004,ChenHongTeo2012}
by $a=c+\epsilon^2$, taking the common $\sqrt{a-c}$ in its potential
to be $\epsilon$. Below, $\mathcal A$, $g$, and $\varphi$ denote the gauge potential,
metric, and scalar (dilaton) field of the dipole parent.
Their expansions about the vacuum ring are
\begin{equation}
\begin{aligned}
\mathcal A(\epsilon)&=\epsilon\mathcal A^{(1)}+O(\epsilon^3),\\
g(\epsilon)&=g_{\rm PS}+O(\epsilon^2),\\
\varphi(\epsilon)&=\epsilon^2\varphi_2+O(\epsilon^4).
\end{aligned}
\label{eq:tangent}
\end{equation}
The second Maxwell field is then
$\Ht=\dd\mathcal A^{(1)}$ locally. The signed convention makes
the two orientations of the dipole amplitude smooth through zero.
Because the metric and dilaton have no first-order perturbation, the
first-order parent gauge equation is
\begin{equation}
\dd\Ht=0,\qquad \dd(*_g\Ht)=0 .
\label{eq:maxwell}
\end{equation}
Although obtained from a first-order expansion, $\Ht$
is an exact source-free Maxwell field on the vacuum
black-ring geometry.

We now define
\begin{equation}
\begin{gathered}
\Phi=1+c-b+bc,\qquad \Psi=1-c-b-bc,\\
Q_1=\frac{\kappa(1+b)\sqrt{2c}}{\sqrt{\Phi\Psi}}>0 .
\end{gathered}
\label{eq:Q}
\end{equation}
The two regular axis potentials differ by $2Q_1\dd\phi$, and hence
define the same global field strength. For a sphere linking the ring,
oriented by $\dd x\wedge\dd\phi$,
\begin{equation}
\frac{1}{4\pi}\int_{S^2}\Ht=Q_1 \,,
\label{eq:flux}
\end{equation}
where reversing the orientation reverses the sign. The nonzero integral prevents
a single smooth global potential, whereas $E$ is globally exact
wherever $\xi^\flat$ is smooth. The two fields therefore carry
different global information, a feature studied in other classical
double-copy settings \cite{AlfonsiWhiteWikeley2020}. For nonzero second spin, $\Ht$ acquires $t$- and $\psi$-components
while retaining its nonzero linking flux.

The asymptotic normalization of time fixes $E$, while the signed parent
amplitude fixes $\Ht$ and $Q_1$. Although the physical parent field
vanishes at zero amplitude, its first-order coefficient survives as a
Maxwell test field; backreaction begins at second order.

\emph{Polarized Weyl identity.---}
For a two-form $F$, we fix the same five-dimensional quadratic map as in
the case of the singly spinning construction~\cite{EassonER2026}:
\begin{equation}
\W_g(F)_{abcd}
=\left[F_{ab}F_{cd}
+\tfrac12(F_{ac}F_{bd}-F_{ad}F_{bc})\right]_{\rm Weyl} .
\label{eq:W}
\end{equation}
The brackets subtract the traces with $g$; the unprojected tensor
already has the Riemann symmetries. Its symmetric polarization is
\begin{equation}
\W_g(E,H)=\frac{\W_g(E+H)-\W_g(E)-\W_g(H)}{2} .
\label{eq:polarization}
\end{equation}
We find for the doubly spinning ring, the Weyl tensor satisfies
\begin{equation}
\boxed{C[g]=A\W_g(E)+B\W_g(E,\Ht)+D\W_g(\Ht).}
\label{eq:copy}
\end{equation}
Writing $\Delta=x-y$, the coefficients are
\begin{align}
A&=-\frac{(1+c^2-b)+c(2+b)x-bcy+bc^2xy}
                  {2c(1+b)\Delta},\label{eq:A}\\
B&=\frac{\sqrt{2bc(1-c^2)}(1-xy)}{(1+b)\Delta},\label{eq:B}\\
D&=\frac{1-c^2-b-bc(x+y)-bc^2xy}{(1+b)\Delta}.
\label{eq:D}
\end{align}
Let $F_1=E$ and $F_2=\Ht$ denote the two Abelian fields, and assemble the weights as
\begin{equation}
\mathbf M=\begin{pmatrix}A&B/2\\ B/2&D\end{pmatrix},
\qquad
C[g]=\sum_{I,J=1}^2 M_{IJ}\W_g(F_I,F_J).
\label{eq:matrix}
\end{equation}

\emph{Matrix zeroth copy and six-dimensional origin.---}
We identify $\mathbf M^{-1}$ as a \emph{matrix zeroth copy}:
its entries are scalar fields sourced by pairwise contractions
of the same two Maxwell fields that reconstruct the Weyl tensor.
In the stationary parent gauge with
$\mathcal A_t^{(1)}\to0$ at infinity,
\begin{equation}
\mathbf S\equiv\mathbf M^{-1}=
\begin{pmatrix}
-(1+\xi^2)&-\xi\!\cdot\!\mathcal A^{(1)}\\
-\xi\!\cdot\!\mathcal A^{(1)}&2\varphi_2/\alpha
\end{pmatrix},
\qquad \alpha=2\sqrt{\frac23}.
\label{eq:scalarmatrix}
\end{equation}
Its entries obey a common equation on the PS geometry,
\begin{equation}
\boxed{\Box_g S_{IJ}=-\frac12 F_{I\,ab}F_J{}^{ab}.}
\label{eq:matrixwave}
\end{equation}
In the six-dimensional Ricci-flat Kaluza--Klein uplift of the
dipole parent, $\mathbf S$ is obtained from the inner products
of two Killing vectors, after subtracting their asymptotic
values and taking an appropriately rescaled vacuum limit.
The six-dimensional Killing identity gives Eq.~\eqref{eq:matrixwave}.
Independently, the PS Weyl identity establishes
$\mathbf S=\mathbf M^{-1}$.
The same two Killing directions generate the Maxwell fields:
$E=\lim_{\epsilon\to0}\dd k_t^\flat$ and
$\Ht=\lim_{\epsilon\to0}\epsilon^{-1}\dd k_w^\flat$,
where $k_t=\partial_t$ and $k_w=\partial_w$ are Killing
vectors of the six-dimensional parent.
The weights are dimensionless and independent of $\kappa$, with positive
real branches for all radicals. The uplift and Gram-matrix limit are given explicitly
in the Supplemental Material~\cite{Supplement}.

The proof checks the Riemann tensor component by component,
using exact polynomial cancellation in a finite algebraic extension. Since
\eqref{eq:W} is tracefree, the identity gives Ricci
flatness and $C=R$. The result holds
at finite ring thickness and both angular momenta.

For the singly spinning ring, reflection of the sphere azimuth
makes $E$ even and $\Ht$ odd, forcing the mixed coefficient
$B$ and the off-diagonal entries of $\mathbf S$ to vanish.
Introducing a second spin breaks this symmetry, and $B$ appears at
linear order in the signed second spin.

\emph{An unavoidable mixed product.---}
A constant recombination of the two Maxwell fields preserves
the source-free Maxwell equations. Writing
$(E,\Ht)^{\mathsf T}=T(\widetilde F_1,\widetilde F_2)^{\mathsf T}$,
the coefficient matrix $\mathbf M$ in Eq.~\eqref{eq:matrix}
transforms as $\mathbf M\mapsto T^{\mathsf T}\mathbf M T$.
Here $\widetilde F_1$ and $\widetilde F_2$ form another
constant basis of the same two-dimensional space of
Maxwell solutions.

For any allowed $b,c$, choose three exterior coordinate
pairs $P_i=(x_i,y_i)$:
\begin{equation}
\begin{gathered}
P_1=\left(0,-\frac{1+c}{2c}\right),\quad
P_2=\left(\frac12,-\frac{1+c}{2c}\right),\\
P_3=\left(0,-\frac{1+2c}{3c}\right).
\end{gathered}
\label{eq:points}
\end{equation}
All three points lie in the open exterior for~$0<c<1$.
Substitution into \eqref{eq:A}--\eqref{eq:D} gives
\begin{equation}
\begin{gathered}
\det\begin{pmatrix}
A(P_1)&B(P_1)&D(P_1)\\
A(P_2)&B(P_2)&D(P_2)\\
A(P_3)&B(P_3)&D(P_3)
\end{pmatrix}\\
=-\frac{bc^3(1-c)^2}{2(1+b)^2(1+c)(1+2c)^2}
\sqrt{2bc(1-c^2)}<0 .
\end{gathered}
\label{eq:witness}
\end{equation}
There is no exceptional allowed parameter pair: every factor in
the magnitude is strictly positive. The three coefficient matrices
are therefore independent. A common invertible basis change preserves
that independence, whereas diagonal matrices have only two independent
entries. Hence, for every allowed $b,c$, no spacetime-independent real
or complex change of basis within $\operatorname{span}\{E,\Ht\}$
eliminates the mixed term throughout the exterior.
The basis may be chosen separately for each
fixed parameter pair; the obstruction still holds.

The Supplement gives our three-point calculations and the proof. 

\emph{Singly spinning limit.---}
At fixed $c$ and $\kappa$, the singly spinning limit gives
\begin{equation}
\begin{gathered}
B\longrightarrow0,\qquad
A\longrightarrow-\frac{1+c^2+2cx}{2c\Delta},\\
D\longrightarrow\frac{1-c^2}{\Delta},\qquad
\Ht\longrightarrow Q_0\,\dd x\wedge\dd\phi,\\
Q_0=\kappa\sqrt{\frac{2c}{1-c^2}} .
\end{gathered}
\label{eq:limit}
\end{equation}
The field limit is finite: its time and $\psi$ potential components
vanish, and the $\phi$ component tends to $Q_0(1+x)$.
The flux field survives while the mixed coefficient vanishes. The
scalar matrix becomes diagonal,
\begin{equation}
\begin{gathered}
S_{12}\longrightarrow0,\qquad
S_{11}\longrightarrow-\frac{2c\Delta}{1+c^2+2cx},\\
S_{22}\longrightarrow\frac{\Delta}{1-c^2}.
\end{gathered}
\label{eq:scalarlimit}
\end{equation}
Where the original fields and metric have their regular limit,
\eqref{eq:copy} becomes the diagonal two-square identity of
Ref.~\cite{EassonER2026}. For the unit-flux field
$H_{\rm ER}=\Ht/Q_0$, its scalar partner is $S_{22}/Q_0^2$, as
in the companion paper \cite{EassonER2026}.

In the companion notation, $\nu=c$,
$\lambda=2c/(1+c^2)$, and $R^2=2\kappa^2(1+c^2)$, with the
same physical angles. If $H_{\rm ER}=\dd x\wedge\dd\phi$ is
the unit-flux representative, homogeneity of the quadratic map
gives
\begin{equation}
\left.D\right|_{b=0}\W_g(Q_0H_{\rm ER})
=\frac{2\kappa^2c}{\Delta}\W_g(H_{\rm ER}).
\label{eq:magneticlimit}
\end{equation}
This reproduces the ER magnetic coefficient in
physical-angle and unit-flux normalization.

\emph{Discussion.---}
The Weyl identity provides a matrix zeroth copy
whose entries satisfy the sourced equations
\eqref{eq:matrixwave}. Thus the gravitational curvature and its
scalar partners are built from the same two Maxwell fields, with
a mixed coupling introduced by the second spin.

The identity constrains the De Smet Weyl polynomial
\cite{DeSmet2002}. For Maxwell
quadratics $q_F$ in the five-dimensional spinor convention of
Ref.~\cite{Godazgar2010}, with normalization specified in the Supplement,
\begin{equation}
P_C=\tfrac32\bigl(Aq_E^2+Bq_Eq_H+Dq_H^2\bigr),\qquad H=\Ht .
\label{eq:spinorcopy}
\end{equation}
Over $\mathbb C$ it factors into two quadratics, giving De Smet type
22 or more special wherever the identity holds. To establish the generic De Smet type, we evaluate the
curvature at an exterior point with exact rational parameters:
$(c,b,x,y,\kappa)=(8/25,256/561,0,-2,1)$.
The resulting Weyl polynomial has two distinct irreducible rank-four
quadratic factors, establishing type 22 at that point and generically.
More special factorizations may occur
elsewhere. The distinct factors exclude a single-Maxwell-square Weyl
representation there, independently of the Maxwell field chosen. The
Supplement gives
the factors and a same-convention ER control. Importantly, the latter disagrees with
the earlier singly rotating type-4 assignment \cite{Godazgar2010};
the companion paper \cite{EassonER2026} gives the detailed
reconciliation. De Smet factorization leaves CMPP alignment unchanged
\cite{ColeyMilsonPravdaPravdova2004,PravdaPravdova2005,MonteiroNicholsonOConnell2019}.
These pointwise factors are algebraic; the obstruction in
\eqref{eq:witness} concerns constant combinations of Maxwell solutions.

The field $\Ht$ carries nonzero flux through the ring's linking
two-cycle, while the signed dipole parent fixes its normalization.
A weighted vacuum limit of the matrix of inner products
of the six-dimensional Killing vectors $\partial_t$ and
$\partial_w$ determines $\mathbf S$ independently of the
Weyl tensor. The second rotation generates a mixed inner
product between these Killing vectors, with
$S_{12}=-\lim_{\epsilon\to0}\widehat G_{tw}/\epsilon$ and
$B=-2S_{12}/\det\mathbf S$.
The second angular momentum thus produces a qualitative
departure from the singly spinning ring, introducing an
irreducible coupling between the stationary and topological
Maxwell sectors. We find the resulting five-dimensional double-copy retains an imprint of the extra Killing symmetry
of its six-dimensional parent.

\acknowledgments
\emph{Acknowledgments---}It is a pleasure to thank  Michael Falato, Cindy Keeler and Tucker Manton for helpful discussions. OpenAI LLMs were used to optimize and debug computational scripts and to assist with copyediting.

\bibliography{references}

\clearpage
\onecolumngrid
\renewcommand{\theequation}{S\arabic{equation}}
\renewcommand{\thepage}{S\arabic{page}}
\renewcommand{\thesection}{\Alph{section}}
\renewcommand{\theHequation}{supplement.\arabic{equation}}
\hypersetup{pageanchor=false}
\setcounter{equation}{0}
\setcounter{section}{0}
\setcounter{page}{1}
\setcounter{secnumdepth}{1}
\begin{center}
{\large\bfseries Supplemental Material for\\[0.4em]
The double copy of the doubly spinning black ring\par}
\vspace{0.6em}
Damien A. Easson
\end{center}

\section{Coordinates, metric functions, and conventions}
\label{sec:geometry}

We adopt coordinates $(t,x,y,\phi,\psi)=(0,1,2,3,4)$; the orbit
order used in the triangular coframe is $(t,\psi,\phi)$. We use metric signature $(-++++)$, both angles have period $2\pi$, and the stationary Killing
vector is $\xi=\partial_t$, with $\xi^2\to-1$ at infinity. We use
\begin{equation}
R^a{}_{bcd}=\partial_c\Gamma^a{}_{db}-\partial_d\Gamma^a{}_{cb}
+\Gamma^a{}_{ce}\Gamma^e{}_{db}-\Gamma^a{}_{de}\Gamma^e{}_{cb},
\qquad R_{abcd}=g_{ae}R^e{}_{bcd},\qquad R_{bd}=g^{ac}R_{abcd}.
\label{eq:curvature}
\end{equation}
The strict vacuum parameter and exterior coordinate ranges are
\begin{equation}
0<c<1,\qquad 0<b<\frac{1-c}{1+c},\qquad\kappa>0,
\qquad -1<x<1,\qquad-1/c<y<-1.
\label{supp:eq:domain}
\end{equation}
Regular-chart qualifications are stated in Sec.~\ref{sec:denominators}.
Here $b$ controls rotation about the $S^2$, and $\kappa$ sets the scale.
Define
\begin{equation}
\begin{gathered}
\Delta=x-y,\quad G(z)=(1-z^2)(1+cz),\quad
\Phi=1+c-b+bc,\quad\Psi=1-c-b-bc,\\
q=\sqrt{2c},\quad p=\sqrt{b(1-c^2)},\quad s=\sqrt{\Phi\Psi}.
\end{gathered}
\label{eq:definitions}
\end{equation}
All radicals are positive in \eqref{supp:eq:domain}.

Introduce
$0<\mu_-<\mu_+<1$ and write
\begin{equation}
c=\frac{\mu_+-\mu_-}{1-\mu_+\mu_-},\qquad
b=\frac{\mu_-(1-\mu_+^2)}{\mu_+(1-\mu_-^2)},\qquad
\lambda_{\rm PS}=\mu_++\mu_-,\qquad
\nu_{\rm PS}=\mu_+\mu_- .
\label{eq:PSparametermap}
\end{equation}
The coordinate transformation in Sec.~6.2 of
Ref.~\cite{ChenHongTeo2012} is
\begin{equation}
x=\frac{\widetilde x+\mu_-}{1+\mu_-\widetilde x},\qquad
y=\frac{\widetilde y+\mu_-}{1+\mu_-\widetilde y}.
\label{eq:PScoordinatemap}
\end{equation}
In these variables the standard horizon polynomial has factors
$(1+\mu_+\widetilde y)(1+\mu_-\widetilde y)$, and
\begin{equation}
\lambda_{\rm PS}^2-4\nu_{\rm PS}=(\mu_+-\mu_-)^2>0,\qquad
1+\nu_{\rm PS}-\lambda_{\rm PS}=(1-\mu_+)(1-\mu_-)>0.
\label{eq:PSmargins}
\end{equation}
Thus $0<\nu_{\rm PS}<1$ and
$2\sqrt{\nu_{\rm PS}}<\lambda_{\rm PS}<1+\nu_{\rm PS}$.
This covers the entire strict domain: for fixed $c$, take
$\mu_+=(c+\mu_-)/(1+c\mu_-)$; then
$b=\mu_-(1-c^2)/[(c+\mu_-)(1+c\mu_-)]$ increases strictly from
$0$ to $(1-c)/(1+c)$ as $\mu_-$ increases from $0$ to $1$.
The map takes $-1/c<y<-1$ to $-1/\mu_+<\widetilde y<-1$,
with $y=-1/c$ mapping to the outer horizon
$\widetilde y_h=-1/\mu_+$. Smoothness and Lorentzian signature of this exterior, and the
horizon identification in the analytic extensions, follow from
the analysis \cite{ChruscielCortierGarcia2010}.
The analysis in Sec.~\ref{sec:denominators} shows that the
Weyl identity extends across loci where the auxiliary coframe
becomes singular, wherever the metric and Maxwell fields
remain smooth in a regular chart.
The PS symbols in \eqref{eq:PSparametermap} are distinct from
the ER parameters used in Sec.~\ref{sec:limit}.

The polynomials are given before specializing the dipole parameter
$a$ to $c$, specifying the parent potential used in
Sec.~\ref{sec:tangent}. Put $\Phi_a=1+a-b+ab$ and
$\Psi_a=1-a-b-ab$, and define
\begin{align}
K_a(x,y)={}&-a^2(1+b)\bigl[bx^2(1+cy)^2+(c+x)^2\bigr]
+\bigl[b(1+cy)-1-cx\bigr]^2+bc^2(1-xy)^2,
\label{eq:Ka}\\
H_a(x,y)={}&-a^2(1+b)\bigl[b(1+cx)(1+cy)xy+(c+x)(c+y)\bigr]
\notag\\
&-a(1+b)(x-y)\bigl[c^2-1+b(1+cx)(1+cy)\bigr]
\notag\\
&+\bigl[b(1+cy)-1-cx\bigr]\bigl[b(1+cx)-1-cy\bigr]
+bc^2(1-xy)^2,
\label{eq:Ha}\\
L_a(x,y)={}&a^2(1+b)\bigl[bx(1+cy)^2+(1+c)(c+x)\bigr]
-a(1-x)\bigl[b^2(1+cy)^2+c^2-1\bigr]
\notag\\
&-\bigl[b(1+cy)-c-1\bigr]\bigl[b(1+cy)-cx-1\bigr]
-bc^2(1-y)(1-xy),
\label{eq:La}\\
J_a(x,y)={}&a^2(1+b)\bigl[bx(1+cx)(1+cy)+(1+c)(c+x)\bigr]
\notag\\
&+a\bigl\{(1-x)\bigl[b(1+cx)+c-1\bigr]\bigl[b(1+cy)+c+1\bigr]
-2bc(1-y)(1+cx)\bigr\}
\notag\\
&-\bigl[b(1+cx)-c-1\bigr]\bigl[b(1+cy)-cx-1\bigr]
-bc^2(1-x)(1-xy),
\label{eq:Ja}\\
\mathsf F_a(x,y)={}&\frac{1-y^2}{\Phi_a\Psi_a}
\Bigl\{bcG(x)\bigl[c(y^2-1)\{a^2(1+b)-b+1\}^2
-4a^2y(1-b^2)(1+cy)\bigr]
\notag\\
&-(1+cy)\Bigl[a^2(1+b)^2
\bigl\{a^2(c+x+bx+bcx^2)^2-(c+x-bx-bcx^2)^2\bigr\}
\notag\\
&\hspace{36mm}-(1-b)^2(1+cx)^2
\bigl\{a^2(1+b)^2-(1-b)^2\bigr\}\Bigr]\Bigr\}.
\label{eq:Fa}
\end{align}
These are the functions of Ref.~\cite{ChenHongTeo2012}; $J_a$ is its
$J_+$, and $\mathsf F_a$ denotes its metric function $F$, not a Maxwell
field. The vacuum identity $K_c=H_c$ holds. Set
\begin{equation}
h=H_c(x,y)=K_c(x,y),\qquad \widehat h=H_c(y,x),\qquad
f=\mathsf F_c(x,y),\qquad \ell=L_c(x,y),\qquad j=J_c(x,y).
\label{eq:vacuumfunctions}
\end{equation}
The three rotation coefficients are
\begin{align}
\omega_1&=\frac{2\kappa c(1+b)(1+y)j}{s\widehat h},
\label{eq:w1}\\
\omega_2&=\frac{2\kappa c^2p(1+b)(1-x^2)}{s\widehat h}
\bigl[(1+cy)(c+bc+by)-c-y\bigr],
\label{eq:w2}\\
\omega_3&=\frac{pc^2(1+b)\Delta(1-x^2)(1-y^2)}{\Phi\Psi f}
\bigl[b(1+cx)(1+cy)(1-b-c^2-bc^2)
-(1-c^2)(1-b+c^2+bc^2)\bigr].
\label{eq:w3}
\end{align}
In terms of
\begin{equation}
P=\frac{2\kappa^2h}{\Delta^2},\qquad
d_0=-\frac{\widehat h}{h},\qquad
d_1=\frac{Pf}{h\widehat h},\qquad
d_2=-\frac{PG(x)G(y)}{f},
\label{eq:diagonalweights}
\end{equation}
the complete metric is
\begin{equation}
\dd s^2=d_0(\dd t+\omega_1\dd\psi+\omega_2\dd\phi)^2
+d_1(\dd\psi+\omega_3\dd\phi)^2+d_2\dd\phi^2
+\frac{P}{\Phi\Psi}\left[\frac{\dd x^2}{G(x)}-\frac{\dd y^2}{G(y)}\right].
\label{eq:metric}
\end{equation}
At $\kappa=1$ this is the metric/coframe used by the source implementation.
The coframe is triangular, not orthonormal; no square roots of its
position-dependent diagonal weights are introduced.

\section{Maxwell fields}
\label{sec:fields}

The stationary potential is
\begin{equation}
\xi^\flat=-\frac{\widehat h}{h}
(\dd t+\omega_1\dd\psi+\omega_2\dd\phi),\qquad E=\dd\xi^\flat.
\label{eq:stationarypotential}
\end{equation}
The first-order dipole gauge potential, defined locally, is
\begin{align}
\mathcal A_t^{(1)}&=-qp\,\frac{c(1+b)(1-xy)\Delta}{h},
\label{eq:At}\\
\mathcal A_\psi^{(1)}&=-\frac{\kappa qp}{s}
\frac{c(1+b)(1+y)}{h}
\bigl[x(1-y)(1+c)\Phi+(1-x)^2(2c+bc+bcy)\bigr],
\label{eq:Apsi}\\
\mathcal A_\phi^{(1)}&=-\frac{\kappa q}{s}
\frac{(1+b)(1+x)\ell}{h},
\qquad
\mathcal A^{(1)}=\mathcal A_t^{(1)}\dd t+
\mathcal A_\psi^{(1)}\dd\psi+\mathcal A_\phi^{(1)}\dd\phi,
\qquad \Ht=\dd\mathcal A^{(1)}\quad\hbox{locally}.
\label{eq:Aphi}
\end{align}
In the CHT time convention, the angular components contain $\kappa$, whereas the time component does not.

The potentially nonzero components of both Maxwell fields
can be written without expanding the polynomials. Define
\begin{equation}
\begin{gathered}
\mathcal D_i[N]=\frac{h\,\partial_iN-N\,\partial_i h}{h^2},\qquad
i\in\{x,y\},\qquad Z=(1+cy)(c+bc+by)-c-y,\\
U=x(1-y)(1+c)\Phi+(1-x)^2(2c+bc+bcy).
\end{gathered}
\label{eq:componentoperator}
\end{equation}
Then
\begin{equation}
\begin{aligned}
E_{it}&=\mathcal D_i[-\widehat h],\\
E_{i\psi}&=\mathcal D_i[-2\kappa c(1+b)(1+y)j/s],\\
E_{i\phi}&=\mathcal D_i[-2\kappa c^2p(1+b)(1-x^2)Z/s],\\
(\Ht)_{it}&=\mathcal D_i[-qpc(1+b)(1-xy)\Delta],\\
(\Ht)_{i\psi}&=\mathcal D_i[-\kappa qpc(1+b)(1+y)U/s],\\
(\Ht)_{i\phi}&=\mathcal D_i[-\kappa q(1+b)(1+x)\ell/s].
\end{aligned}
\label{eq:components}
\end{equation}
Here $i=x$ and $i=y$ each specify three components of each field. Components
with both indices in $(t,\psi,\phi)$ and the $xy$ component vanish;
the rest follow by antisymmetry.

\section{Dipole vacuum limit}
\label{sec:tangent}

The CHT solution has $0\le c\le a<1$ and
$0\le b<(1-a)/(1+a)$ \cite{ChenHongTeo2012}. At a strict vacuum point
of \eqref{supp:eq:domain}, a sufficiently small amplitude stays in this
parameter range. Its gauge potential components are
\begin{align}
\mathcal A_t(a)&=-\sqrt{b(a^2-c^2)(1-a^2)}\,
\frac{c(1+b)(1-xy)\Delta}{K_a},
\label{eq:parentAt}\\
\mathcal A_\psi(a)&=-\sqrt{\frac{2ab(a-c)(1-a^2)}{\Phi_a\Psi_a}}
\frac{\kappa c(1+b)(1+y)}{K_a}
\bigl[x(1-y)(1+c)\Phi_a+(1-x)^2(a+ab+bcy+c)\bigr],
\label{eq:parentApsi}\\
\mathcal A_\phi(a)&=-\sqrt{\frac{2a(a-c)}{\Phi_a\Psi_a}}
\frac{\kappa(1+b)(1+x)L_a}{K_a}.
\label{eq:parentAphi}
\end{align}
Write $a=c+\epsilon^2$ and choose the signed branch
of the dipole gauge potential:
\begin{equation}
\sqrt{b(a^2-c^2)(1-a^2)}\ \mapsto\
\epsilon\sqrt{b(a+c)(1-a^2)} .
\label{eq:signedroot}
\end{equation}
The analogous factors in the angular potentials are treated
with the same sign, giving,
\begin{equation}
\mathcal A(\epsilon)=\epsilon\mathcal A^{(1)}+O(\epsilon^3),\qquad
F(\epsilon)=\epsilon\Ht+O(\epsilon^3).
\label{eq:parentexpansion}
\end{equation}
Evaluating the remaining factors at $a=c$ gives
\eqref{eq:At}--\eqref{eq:Aphi}.
The signed convention fixes the normalization of $\Ht$;
a constant rescaling changes $B,D$ accordingly but leaves
the Weyl identity and constant-basis obstruction unchanged.

The parent metric has the form \eqref{eq:metric} with
\begin{equation}
d_{0,a}=-\frac{H_a(y,x)}{\sqrt[3]{K_a^2H_a}},\quad
P_a=\frac{2\kappa^2\sqrt[3]{K_aH_a^2}}{\Delta^2},\quad
d_{1,a}=\frac{P_a\mathsf F_a}{H_aH_a(y,x)},\quad
d_{2,a}=-\frac{P_aG(x)G(y)}{\mathsf F_a},\quad
\varphi_a=-\sqrt{\frac23}\log\frac{K_a}{H_a},
\label{eq:parentmetric}
\end{equation}
and the base factor is $P_a/(\Phi_a\Psi_a)$. The parent rotation
coefficients are
\begin{align}
\omega_{1,a}&=\sqrt{\frac{2a(a+c)}{\Phi_a\Psi_a}}
\frac{\kappa(1+b)(1+y)J_a}{H_a(y,x)},\notag\\
\omega_{2,a}&=\sqrt{\frac{2ab(a+c)(1-a^2)}{\Phi_a\Psi_a}}
\frac{\kappa c(1+b)(1-x^2)}{H_a(y,x)}
\bigl[(1+cy)(a+ab+by)-c-y\bigr],\notag\\
\omega_{3,a}&=\frac{\sqrt{b(1-a^2)}\,ac(1+b)\Delta(1-x^2)(1-y^2)}
{\Phi_a\Psi_a\mathsf F_a}
\bigl[b(1+cx)(1+cy)(1-b-a^2-a^2b)
-(1-c^2)(1-b+a^2+a^2b)\bigr].
\label{eq:parentrotations}
\end{align}
These quantities depend smoothly on $a$, with the real cube roots
understood on regular source charts. They are even in $\epsilon$.
Since $K_c=H_c$ and $\sqrt[3]{H_c^3}=H_c$,
\begin{equation}
g(\epsilon)=g_{\rm PS}+O(\epsilon^2),\qquad
\varphi(\epsilon)=\epsilon^2\varphi_2+O(\epsilon^4).
\label{eq:evenexpansion}
\end{equation}

\section{Gauge patches and linking flux}
\label{sec:patches}

Take
\begin{equation}
Q_1=\frac{\kappa(1+b)q}{s}.
\label{eq:fluxscale}
\end{equation}
The axis values of \eqref{eq:Aphi} are independent of $y$:
\begin{equation}
\mathcal A_\phi^{(1)}(-1,y)=0,\qquad
\mathcal A_\phi^{(1)}(1,y)=2Q_1.
\label{eq:axisvalues}
\end{equation}
Choose potentials regular near the
two axes,
\begin{equation}
\mathcal A_- =\mathcal A^{(1)},\qquad
\mathcal A_+=\mathcal A^{(1)}-2Q_1\dd\phi,\qquad
\mathcal A_- -\mathcal A_+=2Q_1\dd\phi .
\label{eq:patches}
\end{equation}
After cancellation, the azimuthal coefficients of the two
potentials contain factors $1+x$ and $1-x$, respectively.
Near the corresponding collapsing circles,
$1\pm x=O(\rho^2)$, ensuring smoothness on each axis patch.
The $\psi$ potential also vanishes at $y=-1$.
On the overlap, the potentials differ by the closed form
$2Q_1\dd\phi$. Since $\phi$ is not globally single-valued,
the difference is locally, but not globally, exact.
Both patches therefore define the same global field strength.

For a linking sphere at fixed $t,y,\psi$, oriented by
$\dd x\wedge\dd\phi$,
\begin{equation}
\frac{1}{4\pi}\int_{S^2}\Ht
=\frac{2\pi}{4\pi}\bigl[\mathcal A_\phi^{(1)}(1,y)
-\mathcal A_\phi^{(1)}(-1,y)\bigr]=Q_1.
\label{eq:fluxintegral}
\end{equation}
The opposite orientation gives $-Q_1$. The parent's signed dipole
charge provides the same normalization:
\begin{equation}
\mathcal Q(\epsilon)=
\epsilon\frac{\kappa(1+b)\sqrt{2(c+\epsilon^2)}}
{\sqrt{\Phi_{c+\epsilon^2}\Psi_{c+\epsilon^2}}}
=\epsilon Q_1+O(\epsilon^3).
\label{eq:parentcharge}
\end{equation}
Since $Q_1\ne0$, Stokes' theorem excludes a smooth global potential
on the linking sphere. In contrast, the stationary two-form is exact
where its Killing potential is smooth.

\section{Source-free Maxwell equations}
\label{sec:maxwell}

The parent gauge term is
$-\tfrac14e^{-\alpha\varphi}F_{\mu\nu}F^{\mu\nu}$, with
$\alpha=2\sqrt{2/3}$ \cite{ChenHongTeo2012}. The corresponding equations
are
\begin{equation}
\dd F=0,\qquad \dd\bigl(e^{-\alpha\varphi}*_{g}F\bigr)=0.
\label{eq:parentmaxwell}
\end{equation}
Using \eqref{eq:parentexpansion} and \eqref{eq:evenexpansion}, the
coefficient at order $\epsilon$ is
\begin{equation}
\dd\Ht=0,\qquad
\partial_\mu\left[\sqrt{|\det g_{\rm PS}|}\,
g_{\rm PS}^{\mu\rho}g_{\rm PS}^{\nu\sigma}(\Ht)_{\rho\sigma}\right]=0.
\label{eq:linearmaxwell}
\end{equation}
Corrections to the metric, its inverse and determinant, and the
dilaton all start at order $\epsilon^2$, so none contributes to this
coefficient.

We can verify the Maxwell divergence directly. At $\kappa=1$, let
\begin{equation}
V=\frac{4h}{\Phi\Psi\Delta^4},\qquad V^2=-\det g,\qquad
g^{xx}=\frac{\Phi\Psi G(x)}{P},\qquad
g^{yy}=-\frac{\Phi\Psi G(y)}{P}.
\label{eq:volume}
\end{equation}
Writing the orbit block as $g_{\rm orb}=T^{\mathsf T}dT$, its inverse
is $T^{-1}d^{-1}(T^{-1})^{\mathsf T}$, where
\begin{equation}
T=\begin{pmatrix}1&\omega_1&\omega_2\\0&1&\omega_3\\0&0&1\end{pmatrix},
\qquad
T^{-1}=\begin{pmatrix}1&-\omega_1&\omega_1\omega_3-\omega_2\\
0&1&-\omega_3\\0&0&1\end{pmatrix},
\qquad d=\operatorname{diag}(d_0,d_1,d_2).
\label{eq:inverse}
\end{equation}
For $I,J\in(t,\psi,\phi)$ and $i=x,y$, with no sum on $i$, define
\begin{equation}
\mathcal J^i_I=Vg^{ii}\sum_Jg^{IJ}\partial_i\mathcal A_J^{(1)}.
\qquad
\partial_x\mathcal J^x_I+\partial_y\mathcal J^y_I=0.
\label{eq:currents}
\end{equation}
All three equalities were checked exactly in the eight-dimensional
algebraic field specified in Sec.~\ref{sec:architecture}. The two
base divergences vanish because $(\Ht)_{xy}=0$ and all orbit
derivatives vanish. Either locally constant sign of $V$ gives the
same zero divergence.

For $E$, closure follows from \eqref{eq:stationarypotential}.
The contracted Killing identity with \eqref{eq:curvature} gives
\begin{equation}
\nabla_a E^{ab}=-2R^b{}_a\xi^a=0.
\label{eq:KillingMaxwell}
\end{equation}
Vacuum follows from the known PS solution and independently from
the Riemann identity in Sec.~\ref{sec:identity}. Both are test
fields on the vacuum metric $g_{\rm PS}$.

\section{Scalar matrix}
\label{sec:scalarmatrix}

Let $F_1=E$, $F_2=\Ht$, and write the coefficient matrix of the Letter as
\begin{equation}
\mathbf M=\begin{pmatrix}A&B/2\\B/2&D\end{pmatrix}.
\label{eq:matrixsupp}
\end{equation}
Set
\begin{equation}
\rho=\sqrt{2bc(1-c^2)},\qquad
N_A=1+c^2-b+c(2+b)x-bcy+bc^2xy,
\end{equation}
\begin{equation}
N_D=1-c^2-b-bc(x+y)-bc^2xy.
\end{equation}
The determinant is
\begin{equation}
\det\mathbf M=-\frac{h}{2c(1+b)^2\Delta^2},
\label{eq:matrixdet}
\end{equation}
and
\begin{equation}
\mathbf S\equiv\mathbf M^{-1}
=\frac{(1+b)\Delta}{h}
\begin{pmatrix}
-2cN_D&c\rho(1-xy)\\
c\rho(1-xy)&N_A
\end{pmatrix}.
\label{eq:matrixinverse}
\end{equation}
Using the parent functions defined in
Eqs.~\eqref{eq:Ka}--\eqref{eq:Ha}, the same entries reduce independently to
\begin{equation}
S_{11}=\frac{H_c(y,x)}{h}-1=-(1+\xi^2),\qquad
S_{12}=-\mathcal A_t^{(1)},\qquad
S_{22}=-\left.\partial_a\log\frac{K_a}{H_a}\right|_{a=c}.
\label{eq:Sentries}
\end{equation}

\subsection{Invertibility in the open exterior}
The determinant in \eqref{eq:matrixdet} has no zero in the strict
exterior. A parametrization covering the strict domain is
\begin{equation}
c=\frac{z}{1+z},\quad
b=\frac{w}{(1+2z)(1+w)},\quad
x=\frac{u-1}{u+1},\quad
y=-1-\frac{v}{z(1+v)},\qquad u,v,w,z>0.
\label{eq:positiveparam}
\end{equation}
It covers the full strict domain and gives
\begin{equation}
h=\frac{\mathcal P(u,v,w,z)}
{(u+1)^2(v+1)^2(w+1)^2(z+1)^4(2z+1)^2},
\label{eq:hpositive}
\end{equation}
where $\mathcal P$ has 129 nonzero monomials, all with positive integer
coefficients. Hence $h>0$ and $\det\mathbf M<0$ throughout the open
exterior.

\subsection{Scalar matrix wave equation}
For the stationary entry, $E_{ab}=2\nabla_a\xi_b$ and the vacuum Killing
identity give
\begin{equation}
\Box_g(\xi^2)=2\nabla_a\xi_b\nabla^a\xi^b
=\frac12E_{ab}E^{ab},
\qquad
\Box_gS_{11}=-\frac12E^2.
\label{eq:S11wave}
\end{equation}

For the mixed entry we take the stationary parent gauge,
$\mathcal L_\xi\mathcal A^{(1)}=0$, with
$\mathcal A_t^{(1)}\to0$ at infinity. Cartan's identity gives
\begin{equation}
\dd(\xi\!\cdot\!\mathcal A^{(1)})=-i_\xi\Ht,
\qquad
\nabla_aS_{12}=\xi^b(\Ht)_{ba}.
\end{equation}
Using \eqref{eq:linearmaxwell},
\begin{equation}
\Box_gS_{12}=-\frac12E_{ab}(\Ht)^{ab}.
\label{eq:S12wave}
\end{equation}
The axis patch shift in Eq.~\eqref{eq:patches} is proportional to
$\dd\phi$ and therefore has zero contraction with $\xi$. A stationary
gauge transformation with $\mathcal L_\xi\Lambda$ constant can shift the
potential expression by a constant; the stated asymptotic gauge fixes
that freedom. The matrix $\mathbf S=\mathbf M^{-1}$ itself is gauge
independent in the fixed field-strength basis.

For the lower-right entry the CHT parent action is
\cite{ChenHongTeo2012}
\begin{equation}
I=\frac{1}{16\pi G_5}\int d^5x\sqrt{-g}\left[
R-\frac12(\partial\varphi)^2
-\frac14e^{-\alpha\varphi}\mathcal F_{ab}\mathcal F^{ab}\right],
\qquad \alpha=2\sqrt{\frac23}.
\label{eq:parentaction}
\end{equation}
Its dilaton equation is
\begin{equation}
\Box_g\varphi=-\frac{\alpha}{4}e^{-\alpha\varphi}\mathcal F^2.
\label{eq:dilatoneom}
\end{equation}
Because $K_c=H_c$ and $a=c+\epsilon^2$,
\begin{equation}
\varphi_2=-\sqrt{\frac23}
\left.\partial_a\log\frac{K_a}{H_a}\right|_{a=c},
\qquad
S_{22}=\frac{2}{\alpha}\varphi_2.
\label{eq:S22phi}
\end{equation}
With
\begin{equation}
g(\epsilon)=g_{\rm PS}+O(\epsilon^2),\quad
\mathcal F(\epsilon)=\epsilon\Ht+O(\epsilon^3),\quad
\varphi(\epsilon)=\epsilon^2\varphi_2+O(\epsilon^4),
\end{equation}
the $O(\epsilon^2)$ metric correction acts on the vanishing background
scalar and does not enter. Corrections to the inverse metric in
$\mathcal F^2$ begin at $O(\epsilon^4)$. Equation~\eqref{eq:dilatoneom}
therefore gives
\begin{equation}
\Box_{g_{\rm PS}}\varphi_2=-\frac{\alpha}{4}\Ht^2,
\qquad
\Box_gS_{22}=-\frac12\Ht^2.
\label{eq:S22wave}
\end{equation}
Equations~\eqref{eq:S11wave}, \eqref{eq:S12wave}, and
\eqref{eq:S22wave} combine as
\begin{equation}
\Box_gS_{IJ}=-\frac12F_{I\,ab}F_J{}^{ab}.
\label{eq:matrixwavesupp}
\end{equation}

\subsection{Six-dimensional origin}
The CHT uplift \cite{ChenHongTeo2012} is
\begin{equation}
d\widehat s^2=e^{\varphi/\sqrt6}ds_5^2
+e^{-\sqrt{3/2}\varphi}(\dd w+\mathcal A)^2.
\label{eq:uplift}
\end{equation}
Let $\widehat{\mathbf G}$ be the Gram block of
$(\partial_t,\partial_w)$, set
$D_\epsilon=\operatorname{diag}(1,\epsilon)$ and
$\eta=\operatorname{diag}(-1,1)$. In the same stationary gauge,
\begin{equation}
\widehat G_{tt}=\xi^2+O(\epsilon^2),\qquad
\widehat G_{tw}=\epsilon\mathcal A_t^{(1)}+O(\epsilon^3),\qquad
\widehat G_{ww}=1-\epsilon^2S_{22}+O(\epsilon^4),
\end{equation}
so that
\begin{equation}
\mathbf S=-\lim_{\epsilon\to0}
D_\epsilon^{-1}(\widehat{\mathbf G}-\eta)D_\epsilon^{-1}.
\label{eq:gramjet}
\end{equation}
The subtraction is by the asymptotic matrix $\eta$, not by the actual
vacuum Gram block. For two Killing fields of a Ricci-flat metric,
\begin{equation}
\widehat\Box\,\widehat G_{IJ}
=\frac12(\dd k_I^\flat)_{AB}(\dd k_J^\flat)^{AB}.
\label{eq:gramwave}
\end{equation}
Together with $\dd k_t^\flat\to E$,
$\epsilon^{-1}\dd k_w^\flat\to\Ht$, and
$\widehat\Box f=e^{-\varphi/\sqrt6}\Box_{g(\epsilon)}f$ for a
$w$-independent scalar, the weighted limit of
Eq.~\eqref{eq:gramwave} reproduces \eqref{eq:matrixwavesupp}.
Thus the Killing norm, KK potential, and dilaton entries in five
dimensions are the reduced pieces of one parent Gram object, obtained
without using the PS Weyl coefficients. At the vacuum member
$\varphi=0$, so the conformal factors are unity and the background
operator is the physical PS metric. The independently verified
PS curvature identity establishes $\mathbf S=\mathbf M^{-1}$.

The six-dimensional uplift is understood locally and on
compatible gauge patches, with
$w_+=w_-+2\mathcal Q(\epsilon)\phi$.
A compact $w$ direction with fixed period $L_w$ would require
$4\pi\mathcal Q(\epsilon)/L_w\in\mathbb Z$, which is incompatible
with $\mathcal Q(\epsilon)=\epsilon Q_1+O(\epsilon^3)$
for continuously varying $\epsilon$.
The flux in Eq.~\eqref{eq:fluxintegral} is the first-order
KK magnetic flux of the dipole parent.

\section{Quadratic Weyl map and polarization}
\label{sec:projection}

For any two-form $F$, in five dimensions define
\begin{equation}
T(F)_{abcd}=F_{ab}F_{cd}+\tfrac12(F_{ac}F_{bd}-F_{ad}F_{bc}),\qquad
r_{ab}(F)=\tfrac32F_{ai}g^{ij}F_{bj},\qquad
r(F)=g^{ab}r_{ab}(F)=\tfrac32F_{ab}F^{ab}.
\label{eq:rawmap}
\end{equation}
The notation $r(F)$ for the scalar trace avoids confusion with the
radical $s$ of \eqref{eq:definitions}. The quadratic map is
\begin{equation}
\begin{split}
\W_g(F)_{abcd}={}&T(F)_{abcd}
-\frac13\bigl(g_{ac}r_{bd}-g_{ad}r_{bc}-g_{bc}r_{ad}+g_{bd}r_{ac}\bigr)
+\frac{r(F)}{12}(g_{ac}g_{bd}-g_{ad}g_{bc}).
\end{split}
\label{supp:eq:W}
\end{equation}
The raw tensor has both pair antisymmetries, pair exchange and the
first Bianchi identity, with $g^{ac}T(F)_{abcd}=r_{bd}(F)$.
The trace subtraction gives $g^{ac}\W_g(F)_{abcd}=0$.
These universal algebraic facts, checked by direct contraction, hold
independently of Einstein's equation. The map has the normalization of the singly spinning copy
\cite{EassonER2026}, within the curved-background Weyl-doubling
approach \cite{WeylDoubling2020}.

Polarization means
\begin{equation}
\W_g(F,H)=\tfrac12\bigl[\W_g(F+H)-\W_g(F)-\W_g(H)\bigr],\qquad
\W_g(uF+vH)=u^2\W_g(F)+2uv\W_g(F,H)+v^2\W_g(H).
\label{supp:eq:polarization}
\end{equation}
For direct bilinear evaluation the raw tensor and its trace are
\begin{align}
T(F,H)_{abcd}={}&\tfrac12(F_{ab}H_{cd}+H_{ab}F_{cd})
+\tfrac14(F_{ac}H_{bd}+H_{ac}F_{bd}-F_{ad}H_{bc}-H_{ad}F_{bc}),
\notag\\
r_{ab}(F,H)={}&\tfrac34\bigl(F_{ai}g^{ij}H_{bj}+H_{ai}g^{ij}F_{bj}\bigr),
\qquad r(F,H)=\tfrac32F_{ab}H^{ab}.
\label{eq:bilinear}
\end{align}
The same trace subtraction as in \eqref{supp:eq:W} then gives
$\W_g(F,H)$. In particular, the coefficient $B$ below multiplies
this half-polarized tensor, not twice this tensor.

\section{Weyl identity and coefficient functions}
\label{sec:identity}

On the domain specified in Secs.~\ref{sec:geometry} and
\ref{sec:denominators}, the identity is
\begin{equation}
R[g_{\rm PS}]=A\W_g(E)+B\W_g(E,\Ht)+D\W_g(\Ht),
\label{eq:Ridentity}
\end{equation}
with
\begin{equation}
\begin{aligned}
A&=-\frac{(1+c^2-b)+c(2+b)x-bcy+bc^2xy}{2c(1+b)(x-y)},\\
B&=\frac{\sqrt{2bc(1-c^2)}(1-xy)}{(1+b)(x-y)},\\
D&=\frac{1-c^2-b-bc(x+y)-bc^2xy}{(1+b)(x-y)}.
\end{aligned}
\label{eq:coefficients}
\end{equation}
The symbol $R$ on the left denotes the all-lowered Riemann tensor.
Contracting \eqref{eq:Ridentity}, using the universal tracefreeness
of Sec.~\ref{sec:projection}, gives $R_{ab}=0$. Thus $C_{abcd}=R_{abcd}$,
recovering the Weyl-doubling relation in the Letter.

\section{Algebraic proof}
\label{sec:architecture}

All ingredients depend on $x,y$, with $b,c$ treated as independent
indeterminates. Under simultaneous reversal of the three orbit
coordinates $(t,\phi,\psi)$, the metric is unchanged and both
Maxwell two-forms change sign. Their quadratic tensors are even.
In the coordinate order of Sec.~\ref{sec:geometry}, the bivector
index sets split into
\begin{equation}
\mathcal B_0=(03,04,12,34),\qquad
\mathcal B_1=(01,02,13,14,23,24).
\label{eq:parity}
\end{equation}
The $4\times6$ cross block vanishes structurally. The two symmetric diagonal blocks contain
\begin{equation}
\frac{4(4+1)}2+\frac{6(6+1)}2=31
\label{eq:count31}
\end{equation}
upper-triangular entries, each verified to satisfy
Eq.~\eqref{eq:Ridentity}. Pair exchange supplies the lower
triangles. Although the first Bianchi identity relates these components,
all 31 are checked.

The algebra is performed over
\begin{equation}
\mathbb K=\mathbb Q(b,c,x,y)(q,p,s),\qquad
q^2=2c,\quad p^2=b(1-c^2),\quad s^2=\Phi\Psi,
\qquad (1,q,p,qp,s,qs,ps,qps).
\label{eq:MQ}
\end{equation}
The eight listed elements form a basis: the square classes
are independent, as shown by their odd valuations at
$c$, $b$, and $\Phi$.
Multiplication uses the quadratic relations in
Eq.~\eqref{eq:MQ}, and inversion uses successive conjugation.

The triangular inverse \eqref{eq:inverse} reduces orbit contractions
to three diagonal contractions. In terms of lower Christoffel symbols
$\Gamma_{eab}=\tfrac12(g_{ea,b}+g_{eb,a}-g_{ab,e})$,
the Riemann tensor is evaluated as
\begin{equation}
R_{abcd}=\tfrac12(g_{ad,bc}+g_{bc,ad}-g_{ac,bd}-g_{bd,ac})
+g^{ef}(\Gamma_{ead}\Gamma_{fbc}-\Gamma_{eac}\Gamma_{fbd}).
\label{eq:alllowerR}
\end{equation}
The Weyl terms are evaluated with the direct polarization
\eqref{eq:bilinear}. The componentwise identity follows by exact polynomial cancellation.

\section{Polynomial verification and component counts}
\label{sec:certificates}

For each of the 31 components, the signed terms are
\begin{equation}
R_{abcd},\qquad -A\W_g(E)_{abcd},\qquad
-B\W_g(E,\Ht)_{abcd},\qquad -D\W_g(\Ht)_{abcd}.
\label{eq:signedterms}
\end{equation}
We resolve each into the eight basis slots of \eqref{eq:MQ}. If the
four rational summands in a slot are $n_j/d_j$, a nonzero
polynomial common denominator $L$ gives the identity
\begin{equation}
\sum_{j=1}^4 n_j(L/d_j)=0\quad\hbox{in }\mathbb Q[b,c,x,y].
\label{eq:polynomialsum}
\end{equation}
Every division $L/d_j$ is exact.

Of the $31\times8=248$ component-basis combinations,
217 vanish identically by their radical structure.
The remaining 31 reduce to polynomial identities
of the form \eqref{eq:polynomialsum}.
The 24 parity-forbidden cross-block components vanish
by reflection.

A separate polynomial calculation confirms all 248 cancellations.

\section{Domain and denominator analysis}
\label{sec:denominators}

The component calculation is initially performed where
\begin{equation}
c(1+b)\Delta\Phi\Psi G(x)G(y)\,h\,\widehat h\,f\,K_c(x,y)\ne0.
\label{eq:dense}
\end{equation}
Throughout the strict exterior, all factors except
$\widehat h$ and $f$ are nonzero, with $K_c=h>0$.

The denominator calculation factors all 248 cleared-slot denominators
and all 992 signed-summand denominators over the rational polynomial
ring. In
particular, $\widehat h$ and $f$ cancel from the signed
Riemann and weighted-Weyl summands. The calculation additionally checks
all 25 entries of $gg^{-1}=I$ and verifies that the reduced metric,
inverse, and local-potential denominators use only
\begin{equation}
h,\quad\Phi,\quad\Psi,\quad G(x),\quad G(y),\quad\Delta .
\label{eq:reduceddenominators}
\end{equation}
The metric determinant is
\begin{equation}
\det g=-\frac{16\kappa^8h^2}{\Phi^2\Psi^2\Delta^8}.
\label{eq:detmetric}
\end{equation}
The loci $\widehat h=0$ and $f=0$ arise from the auxiliary
coframe and do not by themselves signal physical singularities.
The identity extends across these loci wherever the metric
and Maxwell fields are smooth in a regular chart.

The axes $x=\pm1$ and $y=-1$, horizon $y=-1/c$,
infinity corner $x=y=-1$, parameter boundaries, and possible zeros
of $h$ are distinguished from the open-chart theorem. The axis
patch statements of Sec.~\ref{sec:patches} apply on regular source
charts. 

\section{Rational-point and numerical controls}
\label{sec:controls}

The generic proof is supplemented by the following pointwise controls.
At
\begin{equation}
(c,b,x,y,\kappa)=\left(\frac8{25},\frac{256}{561},0,-2,1\right),
\qquad(q,p,s)=\left(\frac45,\frac{16}{25},\frac7{25}\right),
\label{eq:rationalpoint}
\end{equation}
all radicals are rational. A separate two-variable Taylor-jet calculation gives $R_{ab}=0$ and zero residual for all 100 ordered
pair-pair entries, with
\begin{equation}
(A,B,D)=\left(-\frac{328929}{653600},
\frac{17952}{102125},\frac{257121}{1021250}\right).
\label{eq:rationalcoefficients}
\end{equation}
At $(c,b,x,y,\kappa)=(1/3,1/5,0,-2,1)$ the coefficients are
\begin{equation}
(A,B,D)=\left(-\frac{47}{72},\frac{\sqrt{15}}{27},\frac{37}{108}\right).
\label{eq:numericalcoefficients}
\end{equation}
The 50-decimal-digit evaluation there has a relative
maximum residual of approximately $8.10\times10^{-49}$.
The family-wide Weyl identity rests on the polynomial identities
in Sec.~\ref{sec:certificates}; the point evaluations provide
independent checks.

At \eqref{eq:rationalpoint}, a pointwise rank-three test checks the
polarized Weyl map on the fixed Maxwell plane. Let $T$ be the $3\times3$ matrix formed from the
components $(0101)$, $(0102)$, and $(0113)$ of the three tensors
$\W_g(E)$, $\W_g(E,\Ht)$, and $\W_g(\Ht)$, in that column order.
All entries of $T$ are rational at this point and have denominators
prime to $101$. Reducing rational entries modulo $101$ gives
\begin{equation}
T\bmod 101=
\begin{pmatrix}
45&11&11\\
97&90&65\\
10&14&27
\end{pmatrix},
\qquad \det T\equiv 67\pmod{101}.
\label{eq:pointwiserank}
\end{equation}
The nonzero determinant establishes linear independence of the
three polarized Weyl tensors at this event. By analyticity,
pointwise independence holds generically in the strict parameter
and exterior domain, though not necessarily at every special locus.
Where the pointwise rank is three, the coefficient triple $(A,B,D)$
of an existing decomposition on $\operatorname{span}\{E,\Ht\}$ is
unique.

\section{De Smet factorization and point verification}
\label{sec:desmet}

We adopt the Lorentzian Clifford convention of Ref.~\cite{Godazgar2010}.
With the usual Pauli matrices, write
\begin{equation}
\begin{gathered}
\gamma_0=\operatorname{diag}(-i\sigma_2,i\sigma_2),\qquad
\gamma_1=\operatorname{diag}(\sigma_1,\sigma_1),\qquad
\gamma_2=\operatorname{diag}(\sigma_3,\sigma_3),\qquad
\gamma_3=\begin{pmatrix}0&i\sigma_2\\-i\sigma_2&0\end{pmatrix},\\
\gamma_5=\gamma_0\gamma_1\gamma_2\gamma_3,\qquad
\Gamma_a=(\gamma_0,\gamma_1,\gamma_2,\gamma_3,i\gamma_5),\qquad
\mathcal C=\gamma_0\gamma_5,\qquad
\{\Gamma_a,\Gamma_b\}=2\eta_{ab},\quad
\eta=\operatorname{diag}(-1,1,1,1,1).
\end{gathered}
\label{eq:Clifford}
\end{equation}
For orthonormal-frame components of a two-form and the Weyl tensor, we
define the symmetric matrices and quartic
\begin{equation}
\begin{gathered}
\mathcal B^{ab}=\tfrac12\mathcal C[\Gamma^a,\Gamma^b],\qquad
f^{(F)}_{AB}=F_{ab}\mathcal B^{ab}_{AB},\qquad
q_F(z)=z^{\mathsf T}f^{(F)}z,\\
\mathcal C_{ABCD}=C_{abcd}\mathcal B^{ab}_{AB}\mathcal B^{cd}_{CD},
\qquad P_C(z)=\mathcal C_{ABCD}z^Az^Bz^Cz^D.
\end{gathered}
\label{eq:spinormap}
\end{equation}
Vector indices are raised with $\eta$; all antisymmetric vector-pair
sums are fully ordered, with no implicit factor $1/2$.
The variables $z=(z_1,z_2,z_3,z_4)$ commute. In particular, $f^{(F)}$
is the unscaled contraction in \eqref{eq:spinormap}, fixing the
normalization of $q_F$ independently of other bispinor conventions.
Direct Clifford contraction of \eqref{supp:eq:W} gives
\begin{equation}
P_{\W(F)}=\tfrac32 q_F^2,\qquad
P_{\W(F,H)}=\tfrac32 q_Fq_H,\qquad
P_C=\tfrac32\left(Aq_E^2+Bq_Eq_H+Dq_H^2\right),\quad H=\Ht .
\label{eq:spinoridentity}
\end{equation}
The normalization can be checked on all 55 symmetric pairs of the
ten two-form basis elements, comparing the 35 quartic coefficients
for each pair. Half-polarization gives the middle identity with
the same factor $3/2$.

The scalar binary quadratic in \eqref{eq:spinoridentity} factors
over $\mathbb C$ at every point. Hence the nonzero Weyl polynomial
is of De Smet type 22 or more special wherever the tensor identity
holds. The calculation below establishes irreducibility and
distinctness at the rational point.
Pointwise combinations of $q_E,q_H$ are algebraic and need not
define Maxwell solutions over a region.
The constant-basis obstruction in Sec.~\ref{sec:obstruction} is a
separate statement about one fixed two-field solution space.

\emph{Doubly spinning point.---}
At \eqref{eq:rationalpoint}, direct evaluation
confirms Ricci flatness, and the metric determinant is
\begin{equation}
\det g=-\frac{2028632401411050241}{374491981806250000}.
\label{eq:pointdetmetric}
\end{equation}
\begin{samepage}
A frame used for the verification has the
coframe $\theta^j$ and weights $w_j$ in Table~\ref{tab:pointframe},
so that $g=\sum_{j=0}^4 w_j\theta^j\theta^j$. Let $\Theta$ be its
matrix in coordinate order $(t,x,y,\phi,\psi)$. Column $a$ of the
frame $V$ is column $j_a$ of $\Theta^{-1}$ divided by
$\sqrt{|w_{j_a}|}$, where $(j_0,j_1,j_2,j_3,j_4)=(3,0,1,2,4)$;
negate the final spatial column to preserve orientation.
Then $V^{\mathsf T}gV=\eta$, with positive real square roots.
\par\end{samepage}

\begin{table}[h]
\centering
\caption{Coframe and metric weights at \eqref{eq:rationalpoint}.}
\label{tab:pointframe}
\begin{tabular}{clc}
\toprule
$j$&$\theta^j$&$w_j$\\\midrule
$0$&$\dd t-\dfrac{99030962176}{60918155921}\dd\phi
-\dfrac{11471260932752}{2010299145393}\dd\psi$
&$\dfrac{8702593703}{9970104697}$\\[3mm]
$1$&$\dd x$&$\dfrac{1424300671}{305978750}$\\[3mm]
$2$&$\dd y$&$\dfrac{1424300671}{330457050}$\\[3mm]
$3$&$\dd\phi$&$-\dfrac{2466613872142497}{24001195814481598}$\\[3mm]
$4$&$\dd\psi-\dfrac{12731614150346752}{12000597907240799}\dd\phi$
&$\dfrac{12000597907240799}{3987093305029450}$\\[2mm]
\bottomrule
\end{tabular}
\end{table}

The curvature quartic is computed directly from
\eqref{eq:spinormap}, before comparison with \eqref{eq:spinoridentity}.
All 35 quartic residual coefficients vanish exactly.

The coefficients
\eqref{eq:rationalcoefficients} have discriminant
\begin{equation}
\mathcal D_*=B^2-4AD=\frac{89730942273}{166872250000}\ne0.
\label{eq:pointdiscriminant}
\end{equation}
Writing $r_*=9970104697$, the factors are
\begin{equation}
\begin{gathered}
\alpha_\pm=\frac{-95744\mp4\sqrt{r_*}}{548215},\qquad
Q_\pm=q_E+\alpha_\pm q_H,\qquad
M_\pm=f^{(E)}+\alpha_\pm f^{(H)},\\
P_C=-\frac{986787}{1307200}\,Q_+Q_- .
\end{gathered}
\label{eq:pointfactors}
\end{equation}
In the specified frame the exact determinants are
\begin{equation}
\begin{gathered}
\det M_\pm=\frac{U_*\pm W_*\sqrt{r_*}}{T_*},\\
U_*=59072019055085305259107370853190011911852155705753600000000,\\
W_*=264488043605925041910737666808762229424809574400000000,\\
T_*=5352699526307521605433635990368155954623607904588587443129.
\end{gathered}
\label{eq:factordeterminants}
\end{equation}
Since $99850^2<r_*<99851^2$, its square root is irrational.
The displayed nonzero integers establish that both
determinants are nonzero. Each $M_\pm$ has rank four, whereas a quadratic product
of two linear forms has symmetric matrix rank at most two. Thus
both $Q_\pm$ are irreducible over $\mathbb C$. The two factor matrices are not proportional, as shown
by the nonzero minor
\begin{equation}
(M_+)_{12}(M_-)_{13}-(M_-)_{12}(M_+)_{13}
=-\frac{70790185423117580474148129954539441356800000}
{23229285964155696595958173248860807516005431796211}
\sqrt{29910314091}\ne0.
\label{eq:factorminor}
\end{equation}
Consequently, the point has De Smet type 22.
The frame components and quadratic-factor coefficients
belong to the multiquadratic number field
\begin{equation}
\mathbb K_*=\mathbb Q\left(i,\sqrt3,\sqrt{17405187406},
\sqrt{19940209394},\sqrt{2244111808654029413},
\sqrt{173531540708725845982}\right).
\label{eq:pointfield}
\end{equation}
The calculation uses square-class arithmetic over a degree-64 field,
with positive roots for positive radicands. The distinct irreducible
factors are established at
\eqref{eq:rationalpoint}; the family-wide quadratic factorization follows
separately from \eqref{eq:spinoridentity}.

\emph{ER convention control and the earlier classification.---}
The same Clifford, charge, curvature-sign and ordered-pair conventions
give, at the rotating ER point
$(\nu,\lambda,C_0,R,x,y)=(1/3,3/5,4/5,1,0,-3/2)$,
\begin{equation}
P_{\rm ER}=-\frac53
\left(q_E+\frac{q_{M_{\rm raw}}}{\sqrt5}\right)
\left(q_E-\frac{q_{M_{\rm raw}}}{\sqrt5}\right),\qquad
M_{\rm raw}=\dd x\wedge\dd\phi_{\rm raw},\qquad
\det M_{{\rm ER},\pm}=\frac{45927}{1250}.
\label{eq:ERspinorcontrol}
\end{equation}
Here $C_0$ is the ER rotation constant and $M_{{\rm ER},\pm}$ are
the matrices of the normalized factors in parentheses. The factors
are nonproportional, consistent with the companion
construction \cite{EassonER2026}. Raw angles are used only for this
control; the physical-angle dictionary of Sec.~\ref{sec:limit}
is unchanged.

This ER control disagrees with Godazgar's earlier type-4 designation
\cite{Godazgar2010}. The companion paper \cite{EassonER2026}
gives the detailed reconciliation, including the defects in the
previously published Weyl polynomial. The ER factors were
computed directly from the curvature, independently of that
expression. De Smet factorization over $\mathbb C$ and CMPP
alignment are distinct classifications.

\section{Obstruction to constant-basis diagonalization}
\label{sec:obstruction}

Fix any parameter pair in \eqref{supp:eq:domain}. We show that no spacetime-independent real or complex change of basis within \(\operatorname{span}\{E,\Ht\}\) eliminates the mixed term throughout the exterior, even when the transformation is allowed to depend on the ring parameters.

For the half-polarization \eqref{supp:eq:polarization}, the coefficient
matrix is
\begin{equation}
M(P)=\begin{pmatrix}A(P)&B(P)/2\\B(P)/2&D(P)\end{pmatrix}.
\label{eq:M}
\end{equation}
Choose
\begin{equation}
P_1=\left(0,-\frac{1+c}{2c}\right),\qquad
P_2=\left(\frac12,-\frac{1+c}{2c}\right),\qquad
P_3=\left(0,-\frac{1+2c}{3c}\right).
\label{eq:genericpoints}
\end{equation}
Table~\ref{tab:exteriormargins} gives their margins to the
exterior boundaries. Every entry is positive for $0<c<1$, so the
points lie in the open exterior and their coefficient denominators
are nonzero. They are distinct: $P_2$ has a different $x$ coordinate,
and $y(P_3)-y(P_1)=(1-c)/(6c)>0$.

\begin{table}[h]
\centering
\caption{Strict exterior margins for the generic witness points.}
\label{tab:exteriormargins}
\begin{tabular}{cccccc}
\toprule
Point&$x+1$&$1-x$&$y+1/c$&$-1-y$&$x-y$\\\midrule
$P_1$&$1$&$1$&$(1-c)/(2c)$&$(1-c)/(2c)$&$(1+c)/(2c)$\\[1mm]
$P_2$&$3/2$&$1/2$&$(1-c)/(2c)$&$(1-c)/(2c)$&$(1+2c)/(2c)$\\[1mm]
$P_3$&$1$&$1$&$2(1-c)/(3c)$&$(1-c)/(3c)$&$(1+2c)/(3c)$\\
\bottomrule
\end{tabular}
\end{table}

Set $\rho=qp=\sqrt{2bc(1-c^2)}>0$. Substitution into the
coefficients \eqref{eq:coefficients} gives Table~\ref{tab:genericwitness}.
Factoring $\rho$ from the middle column gives
\begin{equation}
\det\begin{pmatrix}
A(P_1)&B(P_1)&D(P_1)\\
A(P_2)&B(P_2)&D(P_2)\\
A(P_3)&B(P_3)&D(P_3)
\end{pmatrix}
=-\frac{bc^3(1-c)^2}{2(1+b)^2(1+c)(1+2c)^2}
\sqrt{2bc(1-c^2)}<0.
\label{eq:genericwitnessdet}
\end{equation}
Every factor in the magnitude is strictly positive in
\eqref{supp:eq:domain}.

\begin{table}[h]
\centering
\caption{Witness rows for generic parameters, with the positive radical
$\rho$ factored from the middle column.}
\label{tab:genericwitness}
\begin{tabular}{cccc}
\toprule
Point&$A$&$B/\rho$&$D$\\\midrule
$P_1$&$-\dfrac{bc-b+2c^2+2}{2(1+b)(1+c)}$
&$\dfrac{2c}{(1+b)(1+c)}$
&$\dfrac{c(1-c)[2(1+c)-b]}{(1+b)(1+c)}$\\[3mm]
$P_2$&$\dfrac{bc^2-3bc+2b-4c^2-4c-4}{4(1+b)(1+2c)}$
&$\dfrac{5c+1}{2(1+b)(1+2c)}$
&$\dfrac{c(1-c)[4(1+c)-b(c+2)]}{2(1+b)(1+2c)}$\\[3mm]
$P_3$&$-\dfrac{2bc-2b+3c^2+3}{2(1+b)(1+2c)}$
&$\dfrac{3c}{(1+b)(1+2c)}$
&$\dfrac{c(1-c)[3(1+c)-2b]}{(1+b)(1+2c)}$\\[2mm]
\bottomrule
\end{tabular}
\end{table}

The map $(A,B,D)\mapsto M$ is a linear isomorphism. Thus
$M(P_1),M(P_2),M(P_3)$ span the entire three-dimensional symmetric
matrix space over both $\mathbb R$ and $\mathbb C$.

For every allowed parameter pair, the two Maxwell fields are
linearly independent on a nonempty open set. Indeed, at the axis
limit $x=-1$, $y_0=-(1+c)/(2c)$, their component minor is
\begin{equation}
\begin{split}
&\left[E_{yt}(\Ht)_{x\phi}-(\Ht)_{yt}E_{x\phi}\right]_{x=-1,\,y=y_0}\\
&\quad=-\frac{2\kappa c\sqrt{2c}\,(1-b)(1+b)^2(1-c^2)}
{\sqrt{\Phi\Psi}\,h(-1,y_0)}\ne0 .
\end{split}
\label{eq:axisfieldrank}
\end{equation}
The limit is nonzero throughout \eqref{supp:eq:domain}, since
$h(-1,y_0)=\frac{(1-c)^2}{4}[(1-c^2)(4+b^2)-b(5c^2+3)]>0$.
By continuity the fields are independent at
neighboring interior points. The five-dimensional spinor map
$F\mapsto q_F$ in \eqref{eq:spinormap} is injective. Hence the three
quadratics $q_E^2,q_Eq_H,q_H^2$, and thus
$\W_g(E),\W_g(E,\Ht),\W_g(\Ht)$, are linearly independent there.
The coefficient matrix is consequently unique on an open set for
each allowed parameter pair.

Suppose
$(E,\Ht)^{\mathsf T}=T(\widetilde E,\widetilde H)^{\mathsf T}$
with constant invertible $T$. The coefficient matrix transforms as
$M(P)\mapsto T^{\mathsf T}M(P)T$, preserving the three-dimensional
span of the matrices $M(P_i)$. If the mixed coefficient vanished identically,
all three transformed matrices would be diagonal,
contradicting their linear independence.
Over $\mathbb C$, the quadratic map is bilinear, so the
transformation uses transpose rather than Hermitian conjugation.

For fixed parameters, the coefficient functions are rational
in $x,y$. Thus a transformed mixed coefficient vanishing on
the regular coframe chart vanishes wherever these expressions
are defined, including the three witness points.
Their coefficient denominators are nonzero by
Table~\ref{tab:exteriormargins}.
The result concerns constant basis transformations within
$\operatorname{span}\{E,\Ht\}$; position-dependent transformations
do not act by a common congruence on the three matrices.

\emph{Fixed-parameter controls.---}
At $c=1/3$, $b=1/5$, the points in \eqref{eq:genericpoints} are
$(0,-2),(1/2,-2),(0,-5/3)$. The third row is
\begin{equation}
(A,B,D)(0,-5/3)=\left(-\frac{23}{30},\frac{2\sqrt{15}}{45},\frac25\right),
\qquad \det\bigl[(A(P_i),B(P_i),D(P_i))\bigr]_{i=1}^3
=-\frac{\sqrt{15}}{36450}.
\label{eq:genericfixedcontrol}
\end{equation}
\section{Scale dependence}
\label{sec:scale}

The calculation uses $\kappa=1$. For
\begin{equation}
f_\kappa(\tau,x,y,\phi,\psi)=(\kappa\tau,x,y,\phi,\psi),
\qquad
f_\kappa^*g_\kappa=\kappa^2g_1,\quad
f_\kappa^*E_\kappa=\kappa E_1,\quad
f_\kappa^*H_\kappa=\kappa H_1,
\label{eq:scaling}
\end{equation}
where $H_\kappa$ denotes $\Ht$ at scale $\kappa$.
The all-lowered curvature scales by $\kappa^2$, as does every
quadratic Weyl term. The functions $A,B,D$ are unchanged.
This gives \eqref{eq:Ridentity} at every $\kappa>0$. On two-forms
in five dimensions the Hodge star acquires a factor $\kappa$
under $g\mapsto\kappa^2g$, so the homogeneous Maxwell equation
also extends to all scales.

\section{Singly spinning limit and notation dictionary}
\label{sec:limit}

The coefficient formulas give
\begin{equation}
A\longrightarrow-\frac{1+c^2+2cx}{2c\Delta},\qquad
B\longrightarrow0,\qquad
D\longrightarrow\frac{1-c^2}{\Delta}.
\label{eq:coefficientlimit}
\end{equation}
For $B$, take the positive radical branch with finite $\Delta,1+b$;
then $B^2\to0$. The field calculation
gives
\begin{equation}
\mathcal A_t^{(1)}\to0,\quad
\mathcal A_\psi^{(1)}\to0,\quad
\mathcal A_\phi^{(1)}\to Q_0(1+x),\quad
\Ht\to Q_0 H_{\rm ER},\qquad
Q_0=\kappa\sqrt{\frac{2c}{1-c^2}},\quad
H_{\rm ER}=\dd x\wedge\dd\phi .
\label{eq:fieldlimit}
\end{equation}
The original potential and field components contain no inverse
power of $p$, so their limit is taken directly in these coordinates.

For comparison with the balanced physical-angle convention of
Ref.~\cite{EassonER2026}, the parameter dictionary is
\begin{equation}
\nu=c,\qquad \lambda=\frac{2c}{1+c^2},\qquad
R^2=2\kappa^2(1+c^2),\qquad
\alpha^2=\frac{1-\lambda}{(1-\nu)^2}=\frac1{1+c^2}.
\label{eq:dictionary}
\end{equation}
The coordinates $x,y$ and the physical angles are retained, with
the same orientation of the stationary rotation. These relations
are the vacuum singly spinning specialization of the CHT
coordinates and scale \cite{ChenHongTeo2012}, expressed in the
notation of the companion manuscript. If
\begin{equation}
F_x=1+\lambda x,\qquad
S_e=-\frac{\lambda\Delta}{F_x},\qquad
S_H=\frac{\Delta}{R^2\nu\alpha^2},
\label{eq:ERscalars}
\end{equation}
then
\begin{equation}
A\big|_{b=0}=S_e^{-1},\qquad
D\big|_{b=0}\,Q_0^2=\frac{2\kappa^2c}{\Delta}=S_H^{-1},\qquad
C[g_{\rm ER}]=\frac{\W_{g_{\rm ER}}(E)}{S_e}
+\frac{\W_{g_{\rm ER}}(H_{\rm ER})}{S_H}.
\label{eq:ERcopy}
\end{equation}
The factor $Q_0^2$ accounts for the flux $Q_0$ of the PS tangent
field, compared with unit flux in
\cite{EassonER2026}.
Wherever the metric and fields have a regular limit, the identity
reduces continuously to the balanced ER copy at the $b=0$
boundary of the PS parameter domain. In this normalization,
$B=O(\sqrt b)$, with generically nonanalytic dependence on $b$
at the boundary.

\end{document}